\documentclass[10pt, conference]{IEEEtran}
\IEEEoverridecommandlockouts

\usepackage{easyReview}
\usepackage{theorem} \usepackage{cite} \usepackage{stfloats} \usepackage{epsfig} \usepackage{verbatim}
\usepackage{graphicx} 
\usepackage{amsmath,amssymb,amsfonts}
\usepackage{color}
\usepackage{bm}
\usepackage{xcolor}
\usepackage{algorithm}
\usepackage{algorithmicx}

\usepackage{multirow} 
\usepackage{siunitx}
\usepackage{booktabs}
\usepackage{graphicx} 
\usepackage{booktabs} 
\usepackage{array}
\usepackage{makecell}
\usepackage{cellspace}
\usepackage{tabularx}
\newcolumntype{L}[1]{>{\raggedright\arraybackslash}p{#1}}
\newcolumntype{C}[1]{>{\centering\arraybackslash}p{#1}}
\newcolumntype{R}[1]{>{\raggedleft\arraybackslash}p{#1}}

\usepackage{graphicx}
\usepackage{subfigure}
\usepackage{float}
\usepackage{placeins}
\usepackage{subfig}
\usepackage{overpic}
\usepackage{epstopdf}

\usepackage{CJKutf8}
\usepackage{indentfirst}
\usepackage{amsmath}
\usepackage{enumitem}
\usepackage{textcomp}
\usepackage{gensymb}
\usepackage{makecell}
\usepackage{subfigure}
\usepackage{subcaption}

\usepackage{ifthen}

\usepackage{titlesec}

\newboolean{insertCitation}
\setboolean{insertCitation}{true} 

\newcommand{\conditionalCite}[1]{%
    \ifthenelse{\boolean{insertCitation}}{\cite{#1}}{}%
}

\usepackage{easyReview}
\usepackage{theorem} \usepackage{cite} \usepackage{stfloats} \usepackage{epsfig} \usepackage{verbatim}
\usepackage{graphicx} \usepackage{amssymb} \usepackage{amsmath} \usepackage{color}
\usepackage{bm}
\usepackage{xcolor}
\usepackage{algorithm}
\usepackage{algorithmicx}

\usepackage{multirow} 
\usepackage{siunitx}
\usepackage{booktabs}
\usepackage{graphicx} 
\usepackage{booktabs} 
\usepackage{array}
\usepackage{makecell}
\usepackage{tabularx}

\usepackage{graphicx}
\usepackage{subfigure}
\usepackage{float}
\usepackage{placeins}
\usepackage{subfig}
\usepackage{overpic}
\usepackage{epstopdf}

\usepackage{CJKutf8}
\usepackage{indentfirst}
\usepackage{amsmath}
\usepackage{enumitem}
\usepackage{textcomp}
\usepackage{gensymb}
\usepackage{makecell}
\usepackage{subfigure}
\usepackage{subcaption}

\usepackage{ifthen}

\usepackage{titlesec}

\usepackage{array} 
\usepackage{makecell} 

\usepackage{wrapfig} 

\begin{document}

\thispagestyle{empty}
\pagestyle{empty}

\def\QEDclosed{\mbox{\rule[0pt]{1.3ex}{1.3ex}}}
\def\QEDopen{{\setlength{\fboxsep}{0pt}\setlength{\fboxrule}{0.2pt}\fbox{\rule[0pt]{0pt}{1.3ex}\rule[0pt]{1.3ex}{0pt}}}}
\def\QED{\QEDopen}
\def\proof{}
\def\endproof{\hspace*{\fill}~\QED\par\endtrivlist\unskip}

\title{Fluid Antenna Port Prediction based on Large Language Models}
\author{\IEEEauthorblockN{Yali~Zhang$^*$, Haifan~Yin$^*$, Weidong~Li$^*$, Emil~Bj\"{o}rnson$^\dag$, M\'{e}rouane~Debbah$^\ddag$}
\IEEEauthorblockA{$^*$Huazhong University of Science and Technology, Wuhan, China\\
$^\dag$Division of Communication Systems, KTH Royal Institute of Technology, Stockholm, Sweden\\
$^\ddag$KU 6G Research Center, Department of Computer and Information Engineering, Khalifa University, Abu Dhabi, UAE\\
Email: \{yalizhang, yin, weidongli\}@hust.edu.cn, emilbjo@kth.se, merouane.debbah@ku.ac.ae}
}

\maketitle\thispagestyle{empty}

\begin{abstract} 
This study seeks to utilize large language models (LLMs) to forecast the moving ports of fluid antenna (FA). By repositioning the antenna to the locations identified by our proposed model, we intend to address the mobility challenges faced by user equipment (UE). To the best of our knowledge, this paper introduces, for the first time, the application of LLMs in the prediction of FA ports, presenting a novel model termed Port-LLM. The architecture of our model is based on the pre-trained GPT-2 framework. We designed specialized data preprocessing, input embedding, and output projection modules to effectively bridge the disparities between the wireless communication data and the data format utilized by the pre-trained LLM. Simulation results demonstrate that our model exhibits superior predictive performance under different numbers of base station (BS) antennas and varying UE speeds, indicating strong generalization and robustness ability. Furthermore, the spectral efficiency (SE) attained by our model surpasses that achieved by traditional methods in both medium and high-speed mobile environments.\par
\end{abstract}




\section{Introduction}\label{sec_intro}
In recent decades, the implementation of multiple-input multiple-output (MIMO) technology has significantly enhanced the capacity and reliability of communication systems. Nevertheless, the deployment of conventional fixed antennas constrains the exploitation of spatial freedom. In contrast, the position, referred to as ``port" and configuration of the fluid antenna (FA) can be dynamically modified, allowing it to attain substantial diversity gain across various spatially correlated ports \cite{lq-antenna-2021}.\par

The fluid antenna system (FAS) presents significant advantages and potential applications\cite{zhu2023movable}, including MIMO-FAS for the integration of FAS and MIMO, fluid antenna multiple access (FAMA) for counteracting jamming, and the combination of FAS with reconfigurable intelligent surface (RIS) to mitigate the complexities associated with the optimization of RIS. However, FAS also encounters numerous challenges. A primary difficulty arises from the fact that the channel response of the FA is a highly nonlinear function in relation to the ports, making the identification of the appropriate port for the FA that achieves superior communication performance a significant challenge \cite{FAS-1}.\par

Deep learning (DL) technology has garnered significant interest within the domain of wireless communication and led to notable advancements \cite{superdirective-beamforming-2024}, owing to its robust capabilities in feature extraction and modeling \cite{LLM-Telecom-2024}. Notably, the emergence of large language models (LLMs) in recent years has profoundly transformed the fields of natural language processing (NLP) and artificial intelligence (AI). Currently, researchers have initiated investigations into the application of LLMs within the physical layer of wireless communication networks.\par

The issue of mobility, referred to as the curse of mobility, has consistently been a significant concern within the field of communications. To address this issue, the work in \cite{yin2020addressing} proposed the Vec Prony algorithm and the Prony based angular-delay domain (PAD) algorithm. As research in FAs has progressed, the paper \cite{Li2024TransformingTT} proposed the matrix pencil-based moving port (MPMP) algorithm that utilizes the FA to mitigate mobility issues. However, as previously mentioned, the acquisition of moving ports for the FA presents significant challenges. This situation prompts an inquiry into whether the advanced modeling and generalization abilities of LLMs can be effectively utilized to facilitate the identification of moving ports for the FA, thus aiding in the resolution of the mobility issue.\par

In our paper, to the best of our knowledge, we present the novel application of a large language model in the FA moving port prediction for the first time. Our approach aims to maintain the time-varying channel in a relatively stable state despite the user equipment (UE) mobility by adjusting the position of the FA to the port forecasted by our model. Our proposed method consists of two primary steps for predicting the moving port of the FA: the first involves utilizing the channel tables that encompass historical channel state information (CSI) from all moving ports of the FA to forecast the channel tables for subsequent time periods; the second step entails selecting the port of the FA for the forthcoming time based on the predicted channel tables and the known reference channels that require alignment. Given that the extensive datasets employed for the pre-training of LLMs primarily comprise a variety of textual data, these models inherently lack the capability to interpret wireless communication data. As a result, we have developed specialized modules to align our wireless communication data with natural language data.\par

\emph{Notation:} We use boldface to denote matrices and vectors. $\mathbb{R}^n$ and $\mathbb{C}^n$ denote the spaces of $n$-dimension real and complex numbers, respectively. $\left( \cdot \right) ^T$ represents the transpose. $\text{argmin} \left( \cdot \right) $ refers to the input parameter that minimizes the objective function. $\text{unravel\_index}\left( p,\left( n,m \right) \right) $ denotes the multi-dimensional coordinates associated with the integer $p$ within an $n\times m$ dimensional matrix. $\left| \cdot \right|$ is the absolute value, $\lVert \cdot \rVert $ represents the Euclidean norm, and $\mathbb{E}\left[ \cdot \right] $ represents the expectation operator.

 \section{Channel model}\label{sec_channel_model}
\begin{figure}[h]
    \centering
    \includegraphics[scale=0.55]{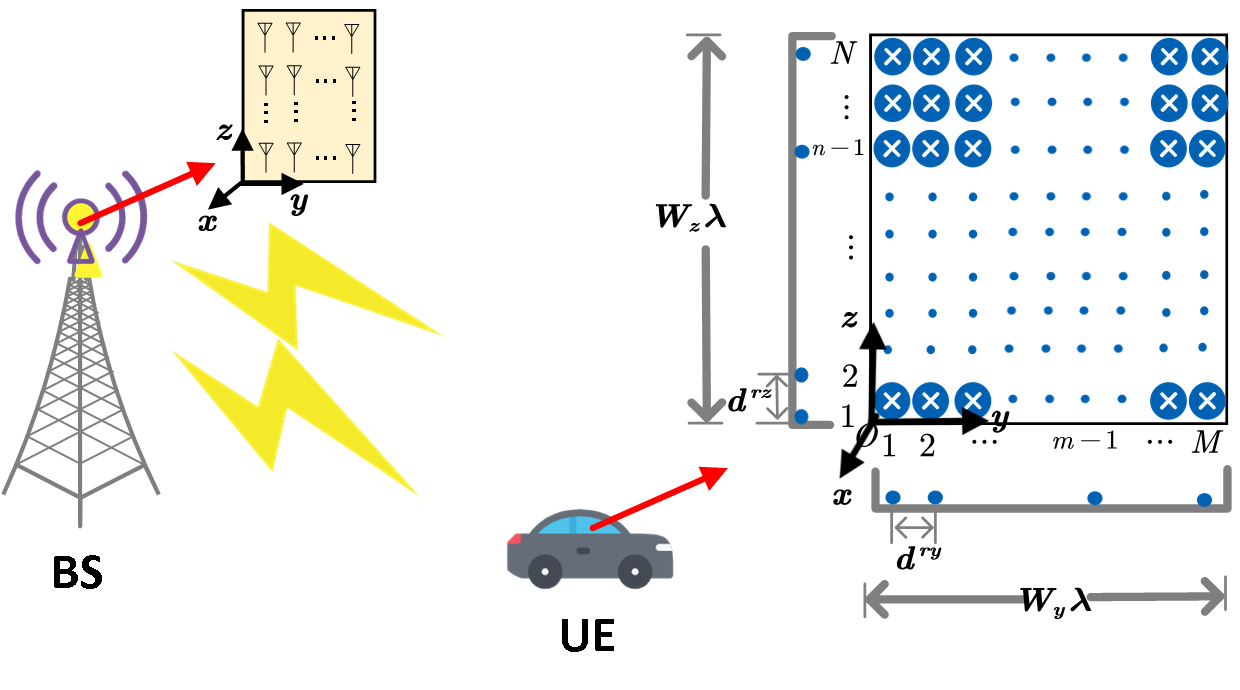}
    \caption{The FA-assisted wireless communication system.}
    \label{fig:BE_UE}
\end{figure}

We consider a system in which the base station (BS) is equipped with an $N_{y}\times N_{z}$ uniform planar array (UPA) situated within the $yOz$ plane, while the UE is equipped with an FA capable of movement within the $yOz$ plane. The size of the movable area of the FA is defined as $W_y\lambda \times W_z\lambda$. The symbol $\lambda =\frac{c}{f_c}$ denotes the wavelength, while $c$ and $f_c$ represent the speed of light and the carrier frequency, respectively. It is assumed that, on the UE side, the quantities of moving antenna ports along the $y$-axis and $z$-axis are denoted by $M$ and $N$, respectively. The schematic representation of our system model is illustrated in Fig.~\ref{fig:BE_UE}.\par

Similar to the clustered delay
line (CDL) channel model used in the study \cite{3GPP2019Study}, we consider a scenario that encompasses one line-of-sight (LoS) path and $P$ non-line-of-sight (NLoS) paths. Therefore, at time $t$, the channel coefficient between all BS-side antennas and the UE-side antenna located at the $\left(n,m\right)$-th port can be represented as
\begin{equation}
    \label{eq:channel_coefficient}
    \begin{split}
        \mathbf{h}_{(n,m)}(t) &= \left[ h_{(1,n,m)}(t), \cdots, h_{(N_t,n,m)}(t) \right]^T \\
        &= \mathbf{A} \mathbf{c}_{(n,m)}(t) \in \mathbb{C}^{N_t\times 1},
    \end{split}
\end{equation}
where $N_t=N_y\times N_z$ is the total number of the BS-side antennas. $h_{\left(i,n,m\right)}\left(t\right)$ denotes the channel coefficient between the $i$-th antenna on the BS side and the UE side antenna at the $\left(n,m\right)$-th port at time $t$. $
\mathbf{A}=\left[ \mathbf{a}\left( \theta _{1}^{\textrm{tx}},\phi _{1}^{\textrm{tx}} \right) ,\mathbf{a}\left( \theta _{2}^{\textrm{tx}},\phi _{2}^{\textrm{tx}} \right) ,\cdots ,\mathbf{a}\left( \theta _{P+1}^{\textrm{tx}},a_{P+1}^{\textrm{tx}} \right) \right] \in \mathbb{C}^{N_t\times \left( P+1 \right)}$ represents the steering vectors of all paths. $\theta_{p}^{\textrm{tx}}$ and $\phi_{p}^{\textrm{tx}}$ denote the elevation angle of departure (EOD) and azimuth angle of departure (AOD) of the $p$-th path, respectively. The 3-D steering vector of the $p$-th path is defined as
\begin{equation}
    \label{eq:steering}
    \mathbf{a}\left( \theta _{p}^{\textrm{tx}},\phi _{p}^{\textrm{tx}} \right) =\mathbf{a}_y\left( \theta _{p}^{\textrm{tx}},\phi _{p}^{\textrm{tx}} \right) \otimes \mathbf{a}_z\left( \theta _{p}^{\textrm{tx}} \right) \in \mathbb{C}^{N_t\times 1},
\end{equation}
where
\begin{equation}
    \label{eq:steering_y}
    \mathbf{a}_y\left( \theta _{p}^{\textrm{tx}},\phi _{p}^{\textrm{tx}} \right) =\left[ 1,\cdots ,e^{j\frac{2\pi}{\lambda}\sin \theta _{p}^{\textrm{tx}}\sin \phi _{p}^{\textrm{tx}}d^{ty}\left( N_y-1 \right)} \right] ^T,
\end{equation}
and
\begin{equation}
    \label{eq:steering_z}
    \mathbf{a}_z\left( \theta _{p}^{\textrm{tx}} \right) =\left[ 1,\cdots ,e^{j\frac{2\pi}{\lambda}\cos \theta _{p}^{\textrm{tx}}d^{tz}\left( N_z-1 \right)} \right] ^T.
\end{equation}\par
Moreover, the vector $\mathbf{c}_{\left( n,m \right)}\left( t \right)$ is given by
\begin{equation}
    \label{eq:8}
    \begin{split}
        \mathbf{c}_{(n,m)}(t) = &\left[ c_{\left(1,n,m\right)} e^{j2\pi w_1 t}, \cdots, \right. \\
        &\left. c_{\left(P+1,n,m\right)} e^{j2\pi w_{P+1} t} \right]^T \in \mathbb{C}^{\left( P+1 \right) \times 1},
    \end{split}
\end{equation}
where $c_{\left( p,n,m \right)}=c_pe^{j\frac{2\pi}{\lambda}\left[ \sin \theta _{p}^{\textrm{rx}}\sin \phi _{p}^{\textrm{rx}}d^{ry}\left( m-1 \right) +\cos \theta _{p}^{\textrm{rx}}d^{rz}\left( n-1 \right) \right]}$ and $c_p=\alpha _p\beta _pe^{j2\pi f\tau _p}$. $\theta_p^{\textrm{rx}}$ and $\phi_p^{\textrm{rx}}$ are the elevation angle of arrival (EOA) and azimuth angle of arrival (AOA) of the $p$-th path, respectively.\par

Assuming $\mathbf{h}_{\left(1,1\right)}\left(t\right)=\left[h_{\left(1,1,1\right)}\left(t\right), \cdots, h_{\left(N_t,1,1\right)}\left(t\right)\right]^T$ is the reference channel. Given the mobility of UE, the channel at any given moment is subject to change relative to the preceding moment, i.e., $\mathbf{h}_{\left(1,1\right)}\left(t\right) \neq \mathbf{h}_{\left(1,1\right)}\left(t+\Delta t\right)$. In order to maintain an approximately constant time-varying channel, it is necessary to adjust the UE-side FA to the appropriate port for the subsequent moment. The following expression can be utilized to determine the moving port $\left(n_{\text{opt}},m_{\text{opt}}\right)$ of FA at time $\left(t+\varDelta t\right)$:
\begin{equation}
\begin{aligned}
\left( n_{\text{opt}},m_{\text{opt}} \right) = 
\text{unravel\_index}\bigg( 
&\text{argmin} 
\bigg( 
\sum_{i=1}^{N_t} 
\Big| \mathbf{S}_i\left( t+\varDelta t \right) - \\
&\mathbf{H}_i\left( t \right) \Big| \bigg)
, \left( N,M \right) \bigg)
\end{aligned},
\end{equation}
where 
\begin{equation}
    \label{eq:port_table_i}
    \mathbf{S}_i\left( t+\varDelta t \right) =\left[ \begin{matrix}{}
    	h_{\left( i,1,1 \right)}\left( t \right)&		\cdots&		h_{\left( i,1,M \right)}\left( t \right)\\
    	\vdots&		\vdots&		\vdots\\
    	h_{\left( i,N,1 \right)}\left( t \right)&		\cdots&		h_{\left( i,N,M \right)}\left( t \right)\\
    \end{matrix} \right] \in \mathbb{C}^{N\times M}
\end{equation}
denotes the channel matrix between the $i$-th antenna at the BS and all ports of the FA at the time $\left(t+\varDelta t\right)$. And the reference channel matrix corresponding to the channel matrix between all ports and the $i$-th antenna at the BS is derived by expanding the $\mathbf{h}_{\left( 1,1 \right)}\left( t \right) \in \mathbb{C}^{N_{t}\times 1}$ as outlined below:
\begin{equation}
    \label{eq:ref_channel_tabel_i}
    \mathbf{H}_{i}\left( t \right) =\left[ \begin{matrix}{}
	h_{\left( i,1,1 \right)}\left( t \right)&		\cdots&		h_{\left( i,1,1 \right)}\left( t \right)\\
	\vdots&		\cdots&		\vdots\\
	h_{\left( i,1,1 \right)}\left( t \right)&		\cdots&		h_{\left( i,1,1 \right)}\left( t \right)\\
\end{matrix} \right] \in \mathbb{C}^{N\times M}.
\end{equation}

The aforementioned formula indicates that when the reference channel is known, determining the moving port of FA at the subsequent moment for maintaining a relatively stable channel hinges on acquiring the channel matrices that connect all antennas on the BS side with all moving ports of the FA on the UE side at that particular moment. For the sake of clarity in the subsequent sections, we will call the channel matrix between the $i$-th antenna on the BS side and all moving ports of FA at a given time as a ``channel table".\par

\section{Proposed scheme}\label{sec_proposed_scheme}
In this section, we introduce our proposed neural network model, referred as Port-LLM, which is designed for the prediction of the moving ports of the FA. Additionally, we provide a comprehensive description of the training methodology employed for our model.\par

The key of forecasting the moving ports of the FA at future time intervals involves the prediction of the channel tables associated with all moving ports of the FA during those intervals. In the subsequent model parameters, in order to simplify the notation, we set the number of antennas on the BS side $N_t$ to 1. Note that the simulation validations are performed in multi-antenna setting. Therefore, the input for our neural network model consists of the channel tables $\mathbf{S}=\left\{ \mathbf{S}_1,\cdots ,\mathbf{S}_T \right\} \in \mathbb{C}^{T\times N\times M}$ relevant to the FA at $T$ historical moments, while the output is the channel tables $\mathbf{\hat{S}}=\left\{ \mathbf{\hat{S}}_1,\cdots ,\mathbf{\hat{S}}_F \right\} \in \mathbb{C}^{F\times N\times M}$ for the subsequent $F$ moments. Subsequently, we proceed to utilize the predicted channel tables $\mathbf{\hat{S}}\in \mathbb{C}^{F\times N\times M}$ and the known reference channel $\mathbf{H}_{\text{ref}} \in \mathbb{C}^{F\times N\times M}$ to obtain the moving ports $\mathbf{P}=\left\{ \mathbf{p}_1,\cdots ,\mathbf{p}_F \right\}\in \mathbb{R}^{F\times 2\times 1}$ for the subsequent $F$ moments.\par 

\subsection{Port-LLM}\label{subsec_network}
\begin{figure*}[t]
    \centering
    \includegraphics[scale=0.50]{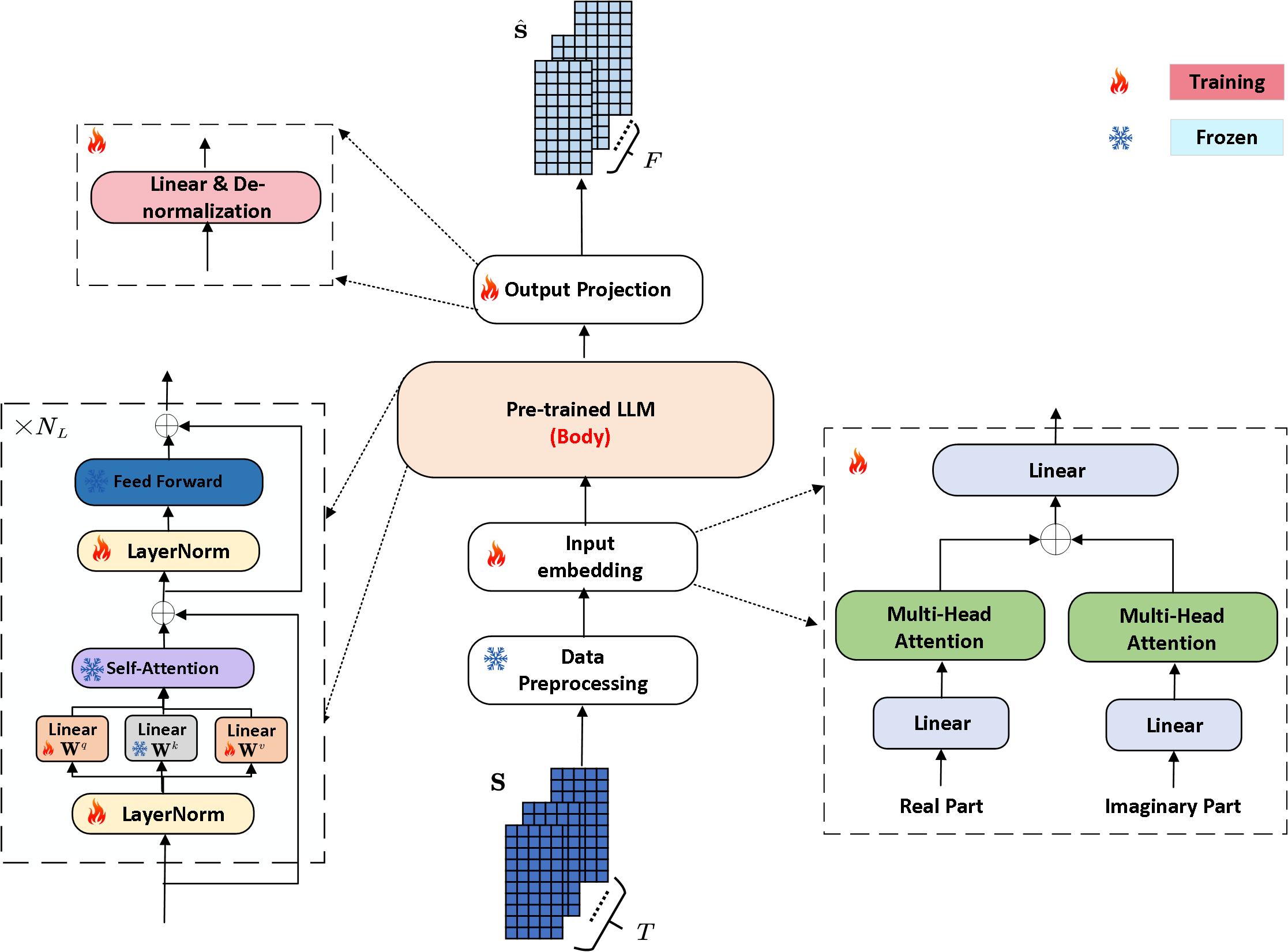}
    \caption{The architecture of our proposed Port-LLM model.}
    \label{fig:PORT-LLM}
\end{figure*}
Our proposed Port-LLM encompasses several components, including data preprocessing, input embedding, backbone network, and output projection modules.\par

\subsubsection{Data Preprocessing}\label{data_preprocessing}
The data preprocessing module initially normalizes the input data $\mathbf{S}\in \mathbb{C}^{T\times N\times M}$, i.e., $\mathbf{\bar{S}}=\frac{\mathbf{S}-\mathbf{\mu }}{\mathbf{\sigma }}$, where $\mathbf{\mu}$ and $\mathbf{\sigma}$ denote the mean and standard deviation of the input data, respectively, and subsequently divides the normalized complex data into its real component $\mathbf{\bar{S}}_r\in \mathbb{R}^{T\times N\times M}$ and imaginary component $\mathbf{\bar{S}}_i\in \mathbb{R}^{T\times N\times M}$, as neural networks typically operate with real-valued data. This data processing step enhances the ability of the subsequent model to extract relevant features from the data and promotes a more rapid convergence of our model.\par

\subsubsection{Input Embedding}\label{input_embedding}
The input embedding module is comprised of the Linear module and the Multi-head attention module \cite{vaswani2017attention}, which primarily serve to extract features from the data along the temporal dimension and to convert the data dimensions into the format required for subsequent input pre-training of the LLM. The Linear module transforms the real part data $\mathbf{\bar{S}}_r\in \mathbb{R}^{T\times N\times M}$ and imaginary part data $\mathbf{\bar{S}}_i\in \mathbb{R}^{T\times N\times M}$ into $\mathbf{\tilde{S}}_r\in \mathbb{R}^{T\times d_{\text{model}}}$ and $\mathbf{\tilde{S}}_i\in \mathbb{R}^{T\times d_{\text{model}}}$, respectively. $d_{\text{model}}$ is the feature dimension of the column vector in the matrix $\mathbf{\tilde{S}}_r$ or $\mathbf{\tilde{S}}_i$. Then the Multi-head attention module is employed to independently extract features from $\mathbf{\tilde{S}}_r$ and $\mathbf{\tilde{S}}_i$. For illustrative purposes, we will focus on the processing of the real component data $\mathbf{\bar{S}}_r$ as a example. In this context, for each head $k \in \left\{ 1,\cdots ,K \right\} $ within the module, we define the query matrix, key matrix, and value matrix as $\mathbf{Q}_{k}^{r}=\mathbf{\tilde{S}}_r\mathbf{W}_{k}^{Q}$, $\mathbf{K}_{k}^{r}=\mathbf{\tilde{S}}_r\mathbf{W}_{k}^{K}$ and $\mathbf{V}_{k}^{r}=\mathbf{\tilde{S}}_r\mathbf{W}_{k}^{V}$, respectively. The reprogramming operation in each attention head is defined as
\begin{equation}
    \label{eq:11}
    \mathbf{S}_{k}^{r}=\text{ATTENTION}\left( \mathbf{Q}_{k}^{r},\mathbf{K}_{k}^{r},\mathbf{V}_{k}^{r} \right) .
\end{equation}\par
Similarly, we also apply the $K$-head multi-head attention module to the imaginary part data $\mathbf{\bar{S}}_i$:
\begin{equation}
    \label{eq:12}
    \mathbf{S}_{k}^{i}=\text{ATTENTION}\left( \mathbf{Q}_{k}^{i},\mathbf{K}_{k}^{i},\mathbf{V}_{k}^{i} \right) ,
\end{equation}
where $\mathbf{S}_{k}^{r}$ and $\mathbf{S}_{k}^{i}\in \mathbb{R}^{T\times d}$. Subsequently, we will integrate the features derived from each head to obtain $\mathbf{S}^r$ and $\mathbf{S}^i\in \mathbb{R}^{T\times d_{\text{model}}}$. In general, we set $d=d_{\text{model}}/K$. Ultimately, we will concatenate these two data to produce the data $\mathbf{X}\in \mathbb{R}^{T\times 2\times d_{\text{model}}}$.\par
\subsubsection{Backbone network}\label{backbone_network}
The backbone network of our proposed model is the loaded pre-trained GPT-2 model \cite{radford2019language}, which has been enhanced through the Low-Rank Adaptation (LoRA) fine-tuning \cite{Hu2021LoRALA}. LoRA fine-tuning significantly reduces the number of parameters that require re-training in the loaded GPT-2 model for this specific task compared to the full fine-tuning technique, while simultaneously enhancing the effective application of the knowledge acquired by the loaded model during its pre-training phase.\par

In particular, we exclusively conduct LoRA fine-tuning and retraining on the $\mathbf{Q}$ and $\mathbf{V}$ computations within the multi-head attention component of the GPT-2 model, while keeping the remaining model parameters frozen. It is assumed that the input data for the module requiring LoRA fine-tuning is denoted by $\mathbf{Z}$. The process of LoRA fine-tuning is outlined as follows:
\begin{equation}
    \label{eq:13}
\mathbf{Q}=\mathbf{W}_Q\mathbf{Z}+\mathbf{B}_Q\mathbf{A}_Q\mathbf{Z}+\mathbf{b}_Q,
\end{equation}
\begin{equation}
    \label{eq:14}
\mathbf{V}=\mathbf{W}_V\mathbf{Z}+\mathbf{B}_V\mathbf{A}_V\mathbf{Z}+\mathbf{b}_V,
\end{equation}
where $\mathbf{W}_Q, \mathbf{W}_V\in \mathbb{R}^{d_m\times d}$ are the weights of the pre-trained GPT-2 model, which remain constant and are not subject to gradient updates throughout the training process. $\mathbf{b}_Q$ and $\mathbf{b}_V$ are the biases of the loaded model, which also remain fixed. $\mathbf{B}_Q, \mathbf{B}_V\in \mathbb{R}^{d_m\times r}$ and $\mathbf{A}_Q, \mathbf{A}_V\in \mathbb{R}^{r\times d}$ are learnable parameters. Furthermore, $r\ll \min \left( d_m,d \right) $, resulting in negligible additional inference delay during model prediction when employing LoRA fine-tuning. Moreover, we employ random Gaussian initialization for parameter $\mathbf{A}_Q$ and $\mathbf{A}_V$, and zero initialization for parameter $\mathbf{B}_Q$ and $\mathbf{B}_V$ prior to the commencement of our model training. \par
Generally, the data $\mathbf{X}\in \mathbb{R}^{T\times 2\times d_{\text{model}}}$ is processed through a linear layer to modify its dimensionality to $\mathbf{\tilde{X}}\in \mathbb{R}^{F\times d_{\text{model}}}$ prior to being fed into the backbone network. Subsequently, the data is integrated into the backbone network, where the following procedure occurs:
\begin{equation}
    \label{eq:16}
    \mathbf{X}_{\text{LLM}}=\text{LLM}_{\text{LoRA}}\left( \mathbf{\tilde{X}} \right) \in \mathbb{R}^{F\times d_{\text{model}}},
\end{equation}
where $\text{LLM}_{\text{LoRA}}\left( \cdot \right) $ represents the LLM-based backbone network that has been fine-tuned by LoRA.\par

\subsubsection{Output Projection}\label{output_projection}
In the output layer, a Linear (or fully connected: FC) layer is employed in conjunction with a rearrange-tensor operation to derive the final output of the model by transforming the dimensions of $\mathbf{X}_{\text{LLM}}$ in the following manner:
\begin{equation}
    \label{eq:17}
    \mathbf{Y}=\text{rearrange}\left( \text{FC}\left( \mathbf{X}_{\text{LLM}} \right) \right) \in \mathbb{R}^{F\times 2\times N\times M}.
\end{equation}\par
Subsequently, execute the denormalization process
\begin{equation}
    \label{eq:18}
    \mathbf{\hat{Y}}=\mathbf{\sigma Y}+\mathbf{\mu },
\end{equation}
where $\mathbf{\hat{Y}}\in \mathbb{R}^{F\times 2\times N\times M}$. The second dimension corresponds to the real and imaginary components of the prediction channel tables, respectively. Additionally, the final output data 
$\mathbf{\hat{S}}\in \mathbb{C}^{F\times N\times M}$ is obtained as follows:
\begin{equation}
    \label{eq:output-data}
    \mathbf{\hat{S}}=\mathbf{\hat{Y}}\left[ :,0,:,: \right] +j\mathbf{\hat{Y}}\left[ :,1,:,: \right]. 
\end{equation}\par

\subsubsection{Moving Port prediction}\label{predicted-ports}
Upon acquiring the channel tables $\mathbf{\hat{S}}=\left\{ \mathbf{\hat{S}}_1,\mathbf{\hat{S}}_2,\cdots ,\mathbf{\hat{S}}_F \right\} \in \mathbb{C}^{F\times N\times M}$ for the subsequent $F$ moments, we employ the predicted channel tables in conjunction with the associated known reference channel $\mathbf{H}_{\text{ref}}=\left\{ \mathbf{H}_{\text{ref}_1},\mathbf{H}_{\text{ref}_2},\cdots ,\mathbf{H}_{\text{ref}_F} \right\} \in \mathbb{C}^{F\times N\times M} $ to derive the final predicted moving ports of the FA for the forthcoming $F$ moments, utilizing the following formula:
\begin{equation}
    \label{eq:optimal-port}
    \mathbf{p}_i=\text{unravel\_index}\left( \text{argmin}\left( \left| \mathbf{\hat{S}}_i-\mathbf{H}_{\text{ref}_i} \right| \right) ,\left( N,M \right) \right)  ,
\end{equation}
where $\mathbf{P}=\left[ \mathbf{p}_1,\cdots ,\mathbf{p}_F \right] \in \mathbb{R}^{F\times 2\times 1}, \mathbf{p}_i=\left[ n_i,m_i \right] ^T\in \mathbb{R}^{2\times 1} , 1\le i\le F, 1\le n_i\le N,1\le m_i\le M$ denotes the predicted moving port of FA at the subsequent $i$-th moment. Here, $n_i$ and $m_i$ correspond to the port indices associated with the predicted moving port of FA along the $z$-axis and $y$-axis, respectively, at the subsequent $i$-th time instance.\par

\subsection{Training of nerual network}\label{subsec_training_ntework}
Our proposed method for predicting the moving port of the FA is a two-step prediction method. Initially, the channel table is forecasted by our proposed model. Subsequently, the moving port of the FA is determined based on the predicted channel table and the known reference channel.\par


Our proposed model is initially trained on channel table datasets and then applied for testing. During the model training process, the objective function is the normalized mean square error (NMSE) between the channel tables $\mathbf{\hat{S}}$ predicted by our model for the future $F$ moments and the actual channel tables $\mathbf{S}$ for these $F$ moments ,as defined in Eq. (\ref{eq:nmse_train}). Concurrently, the NMSE is computed between the channel associated with the port predicted by our model and the reference channel, serving to validate the accuracy of the predicted ports during the training phase, as shown in Eq. (\ref{eq:nmse_val}).\par
\begin{equation}
    \label{eq:nmse_train}
    \mathcal{L}_{\text{Port}-\text{LLM}}=\frac{\lVert \mathbf{S}-\mathbf{\hat{S}} \rVert ^2}{\lVert \mathbf{S} \rVert ^2},
\end{equation}
\begin{equation}
    \label{eq:nmse_val}
    \mathcal{L}_{\text{validate}}=\frac{\lVert \mathbf{h}_{\text{ref}}-\mathbf{h} \rVert ^2}{\lVert \mathbf{h}_{\text{ref}} \rVert ^2}.
\end{equation}

\section{Numerical Results}\label{sec:numericalResult}
In this section, we present the specific parameters utilized in the training of our model, as well as the performance metrics employed for evaluation. Furthermore, we analyze the prediction performance of our model for FA moving ports and conduct the comparative analysis with established methodologies for addressing FA moving ports.\par
\addtolength{\topmargin}{0.05in}
\begin{table}[t]
    \centering
    \caption{The Main Simulation Parameters}
    \label{tab:simulation-parameters}
    \begin{tabularx}{\linewidth}{|>{\centering\arraybackslash}c|>{\centering\arraybackslash}X|}
        \hline
        Channel Model & CDL-D \\ \hline
        Carrier Frequency (GHz)   & 39   \\ \hline
        CSI Delay (ms) & 4 
        \\ \hline
        Delay Spread (ns) & 616 
        \\ \hline
        Sampling Time  & $T_0=5, 6, 10$
        \\ \hline
        UE FA Configuration & 
        \begin{minipage}{\linewidth}
        \centering
        \footnotesize
        $\left(W_y,W_z\right)=\left(10,20\right)$,\\
        $\left(M,N\right)=\left(100,50\right)$,\\
        $\left(\rho_y,\rho_z\right)=\left(5,5\right)$
        \end{minipage}
        \\ \hline
        RMS Angular Spreads       & 
        \begin{minipage}{\linewidth}
        \centering
        \footnotesize
        $\left[31^\degree, 149^\degree, 150^\degree, 30^\degree\right]$, $\left[-38^\degree, 218^\degree, 227^\degree, -47^\degree\right]$, \\
        $\left[1^\degree, 179^\degree, 99^\degree, 81^\degree\right]$, $\left[10^\degree, 170^\degree, 36^\degree, 144^\degree\right]$, \\
        $\left[149^\degree, 31^\degree, 53^\degree, 127^\degree\right]$, $\left[129^\degree, 51\degree, 71^\degree, 109^\degree\right]$, \\
        $\left[-15^\degree, 195^\degree, 210^\degree, -30^\degree\right]$, $\left[199^\degree, -19^\degree, 212^\degree, -32^\degree\right]$, \\
        $\left[-43^\degree, 223^\degree, 76^\degree, 104^\degree\right]$, $\left[7^\degree, 173^\degree, 23^\degree, 157^\degree\right]$
        \end{minipage}
        \\ \hline
    \end{tabularx}
\end{table}

\subsection{Training parameters}
To mitigate the computational demands associated with the training of our model, we adopt a single-input single-output (SISO) system for the acquisition of training datasets. We employ the CDL channel model as defined by the 3rd generation partnership project (3GPP) \cite{3GPP2019Study}. The channel model includes 37 paths, which comprise a LoS path and 36 NLoS paths. The velocity of UEs ranges from 90 km/h to 150 km/h. Each slot contains 14 OFDM symbols, and the duration of a slot is 1 ms. Each group of 50 time slots has a sampling time. To enhance the quantity and diversity of the training dataset, we conducted simulations of communication channels for 10 UEs positioned in various orientations. During the simulation of each UE, we randomly selected distinct sampling time $T_0$ within the designated time slot. Detailed parameter information is shown in Table \ref{tab:simulation-parameters}. A total of 54,300 samples were collected, with 75\% of the dataset allocated for the training set and the remaining 25\% designated for the test set. \par

In the training of our model, the forecasting period for the FA moving ports is established at $F=8$. Concurrently, the duration for the employed channel tables is also designated as $T=8$. As previously indicated, the dimension of the FA moving port table is set as $N\times M=100\times 50$. We employ the smallest version of the GPT-2 model with 768 feature dimensions and utilize only the initial $N_L=6$ layers of the pre-trained GPT-2 architecture. For the LoRA fine-tuning operation, the value of the reduction ratio is set as $r=4$. Furthermore, the number of heads of the multi-head attention module employed in our model is $K=8$. And the dimension is set as $d_{\text{model}}=768$ in the multi-head attention. The Adam algorithm and warm-up-aided cosine LR scheduler are employed to update the parameters. The total number of training epochs for our model is 500.\par

\subsection{Performance metrics}
To assess the performance of our proposed model, the $\text{NMSE}_{\text{t}}$ is employed to quantify the disparity between the actual channel tables $\mathbf{S}$ and the predicted $\mathbf{\hat{S}}$, as described in Eq. (\ref{eq:nmse}).\par
\begin{equation}
    \label{eq:nmse}
    \text{NMSE}_{\text{t}}=10\log _{10}\left\{ \mathbb{E}\left[ \frac{\lVert \mathbf{\hat{S}}-\mathbf{S} \rVert ^2}{\lVert \mathbf{S} \rVert ^2} \right] \right\} .
\end{equation}\par
Furthermore, to enhance the validation of the predicted moving port of FA at the subsequent time point, we computed the NMSE between the channel associated with the predicted port and the reference channel. This metric is referred to as $\text{NMSE}_{\text{v}}$, defined as
\begin{equation}
    \label{eq:validate_nmse}
    \text{NMSE}_{\text{v}}=10\log _{10}\left\{ \mathbb{E}\left[ \frac{\lVert \mathbf{h}-\mathbf{h}_{\text{ref}} \rVert ^2}{\lVert \mathbf{h}_{\text{ref}} \rVert ^2} \right] \right\}.
\end{equation}\par
Additionally, we conducted a comparative analysis of the spectral efficiency (SE) derived from our model against the SE achieved through the Vec Prony algorithm and the MPMP algorithm, respectively. It is computed as
\begin{equation}
    \label{eq:se}
    \text{SE}=\sum_{u=1}^{N_{\text{UE}}}{\mathbb{E}}\left\{ \log _2\left( 1+\text{SINR}_u \right) \right\},
\end{equation}
where $N_{\text{UE}}$ is the number of UEs, $\text{SINR}_u$ denotes signal-to-interference-and-noise radio of the $u$-th UE.\par

\subsection{Performance evaluation}
In practical applications addressing mobility challenges, it is common for BS antennas to be configured as multi-antenna systems. In order to the effectiveness of our model in the MISO system, we investigate the prediction performance of our model with respect to the FA moving ports for different numbers of antennas on the BS side and for varying UE movement speeds. Specifically, our evaluation considers cases involving $2\times 8$, $8\times 8$, and $32\times 8$ antennas at the BS, respectively, and UE speeds of 90 km/h, 120 km/h, and 150 km/h, respectively. It is noteworthy that when evaluating the performance of our proposed models in MISO scenarios, we directly employ the previously trained model under SISO conditions without retraining for different MISO configurations.\par

\begin{figure*}[t]
\centering
\subfigure[]
{
    \begin{minipage}[b]{.31\linewidth}
        \centering
        \includegraphics[scale=0.37]{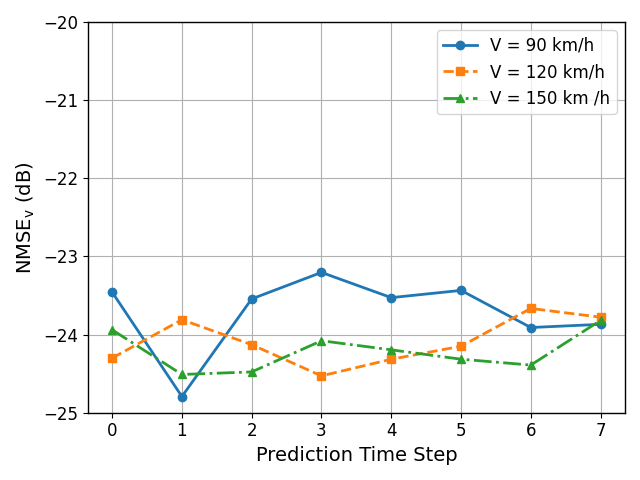}
    \end{minipage}
}
\subfigure[]
{
 	\begin{minipage}[b]{.31\linewidth}
        \centering
        \includegraphics[scale=0.37]{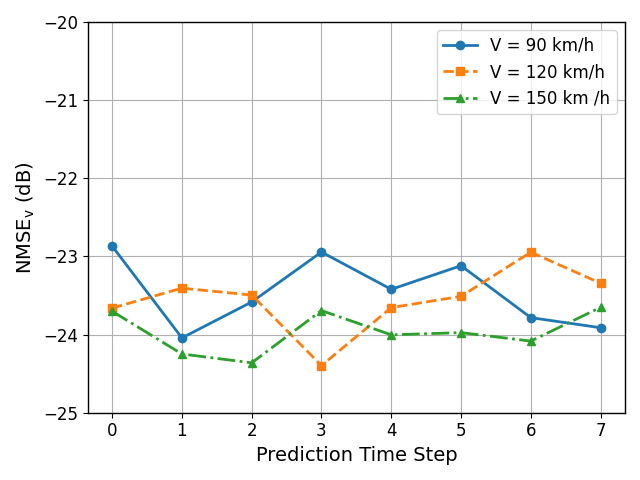}
    \end{minipage}
}
\subfigure[]
{
 	\begin{minipage}[b]{.31\linewidth}
        \centering
        \includegraphics[scale=0.37]{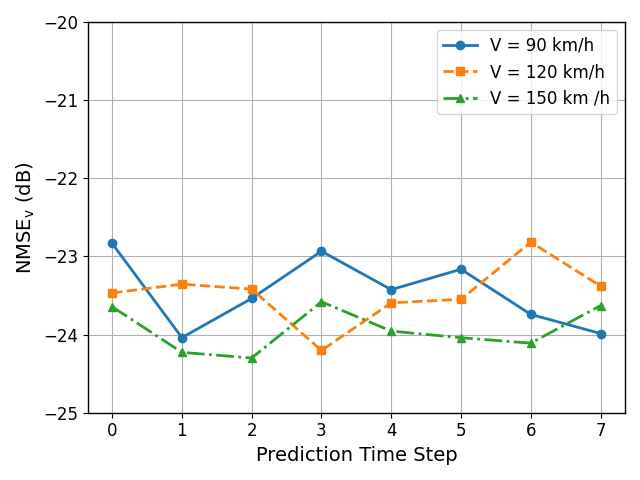}
    \end{minipage}
}
\caption{The predictive performance of our model under different number of antennas on the BS and various UE velocities. (a) The BS has $2\times 8$ antennas; (b) The BS has $8\times 8$ antennas; (c) The BS has $32\times 8$ antennas.}
\label{fig:antenna_veocity}
\end{figure*}

\begin{figure*}[t]
\centering
\subfigure[]
{
    \begin{minipage}[b]{.31\linewidth}
        \centering
        \includegraphics[scale=0.44]{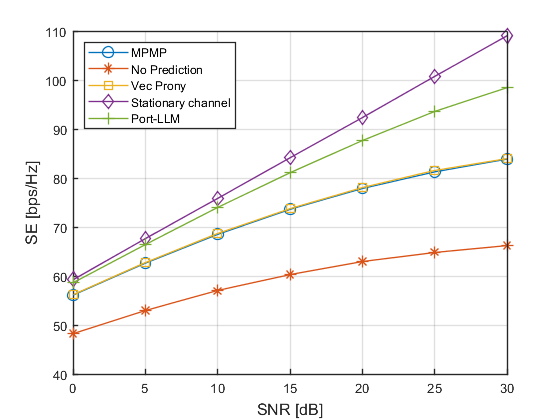}
    \end{minipage}
}
\subfigure[]
{
 	\begin{minipage}[b]{.31\linewidth}
        \centering
        \includegraphics[scale=0.44]{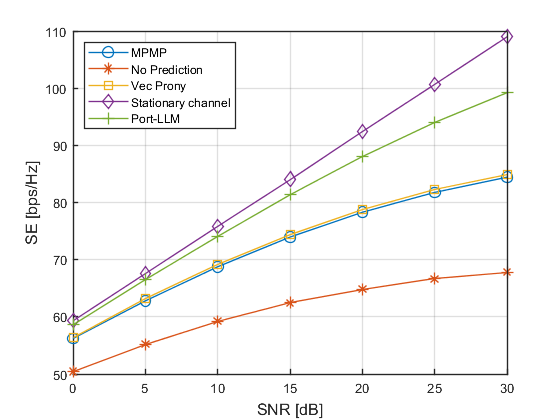}
    \end{minipage}
}
\subfigure[]
{
 	\begin{minipage}[b]{.31\linewidth}
        \centering
        \includegraphics[scale=0.44]{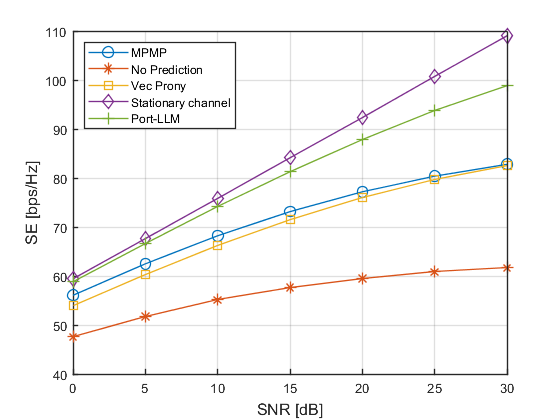}
    \end{minipage}
}
\caption{When the BS has $8\times 8$ antennas, the SE versus SNR. (a) The UE test velocity is 90 km/h; (b) The UE test velocity is 120 km/h; (c) The UE test velocity is 150 km/h.}
\label{fig:se}
\end{figure*}

As illustrated in Fig. \ref{fig:antenna_veocity}, our trained model demonstrates good performance in predicting the moving port for FA across varying numbers of antennas at the BS and differing speeds of UE movement. This finding indicates that our model possesses strong generalization capabilities and robustness under different numbers of BS antennas and medium-to-high mobility speeds of UE. Similarly, with the BS equipped with an $8\times 8$ antenna configuration and varying UE test speeds, we perform a comparative analysis of the SE achieved by our proposed model against the SE derived from both the Vec Prony \cite{yin2020addressing} algorithm and the MPMP \cite{Li2024TransformingTT} algorithm. Additionally, we also evaluated the SE under the idealized scenario (``Stationary channel") and the absence of channel prediction (``No Prediction"). As illustrated in Fig. \ref{fig:se}, optimal performance is observed under ``Stationary channel" condition. Conversely, performance is significantly diminished in the ``No Prediction" condition. Moreover, the SE obtained from the ports predicted by our model surpasses that obtained from both the traditional methods at medium and high speeds.\par

\section{Conclusions}\label{conclusion}
The port prediction method for FA, utilizing Port-LLM as proposed in our paper, demonstrates a notable enhancement in performance relative to conventional approaches. Our model also reveals strong generalization ability and robustness under diverse BS antenna configurations and varying UE movement velocities.\par

\section{Acknowledgment}\label{acknowledgment}
This work was supported by the Fundamental Research Funds for the Central Universities and the National Natural Science Foundation of China under Grant 62071191. E. Bj\"{o}rnson was supported by the Grant 2022-04222 from the Swedish Research Council.\par

\bibliographystyle{IEEEtran}
\bibliography{ref}

\begin{thebibliography}{10}
\providecommand{\url}[1]{#1}
\csname url@samestyle\endcsname
\providecommand{\newblock}{\relax}
\providecommand{\bibinfo}[2]{#2}
\providecommand{\BIBentrySTDinterwordspacing}{\spaceskip=0pt\relax}
\providecommand{\BIBentryALTinterwordstretchfactor}{4}
\providecommand{\BIBentryALTinterwordspacing}{\spaceskip=\fontdimen2\font plus
\BIBentryALTinterwordstretchfactor\fontdimen3\font minus \fontdimen4\font\relax}
\providecommand{\BIBforeignlanguage}[2]{{%
\expandafter\ifx\csname l@#1\endcsname\relax
\typeout{** WARNING: IEEEtran.bst: No hyphenation pattern has been}%
\typeout{** loaded for the language `#1'. Using the pattern for}%
\typeout{** the default language instead.}%
\else
\language=\csname l@#1\endcsname
\fi
#2}}
\providecommand{\BIBdecl}{\relax}
\BIBdecl

\bibitem{lq-antenna-2021}
Y.~Huang, L.~Xing, C.~Song, S.~Wang, and F.~Elhouni, ``{Liquid Antennas: Past, Present and Future},'' \emph{IEEE Open J. Antennas Propag.}, vol.~2, pp. 473--487, 2021.

\bibitem{zhu2023movable}
L.~Zhu, W.~Ma, and R.~Zhang, ``{Movable antennas for wireless communication: Opportunities and challenges},'' \emph{IEEE Commun. Mag.}, vol.~62, no.~10, pp. 114--120, 2023.

\bibitem{FAS-1}
K.-K. Wong, W.~K. New, X.~Hao, K.-F. Tong, and C.-B. Chae, ``{Fluid Antenna System—Part I: Preliminaries},'' \emph{IEEE Commun. Lett.}, vol.~27, no.~8, pp. 1919--1923, 2023.

\bibitem{superdirective-beamforming-2024}
Y.~Zhang, H.~Yin, and L.~Han, ``{A Superdirective Beamforming Approach based on MultiTransUNet-GAN},'' \emph{IEEE Trans. Commun.}, vol.~73, no.~3, pp. 1975--1986, 2025.

\bibitem{LLM-Telecom-2024}
A.~Maatouk, N.~Piovesan, F.~Ayed, A.~D. Domenico, and M.~Debbah, ``{Large Language Models for Telecom: Forthcoming Impact on the Industry},'' \emph{IEEE Commun. Mag.}, pp. 1--7, 2024.

\bibitem{yin2020addressing}
H.~Yin, H.~Wang, Y.~Liu, and D.~Gesbert, ``{Addressing the curse of mobility in massive MIMO with prony-based angular-delay domain channel predictions},'' \emph{IEEE J. Sel. Areas Commun.}, vol.~38, no.~12, pp. 2903--2917, 2020.

\bibitem{Li2024TransformingTT}
W.~Li, H.~Yin, F.~Fu, Y.~Cao, and M.~Debbah, ``{Transforming Time-Varying to Static Channels: The Power of Fluid Antenna Mobility},'' \emph{IEEE Trans. Wireless Commun.}, pp. 1--1, 2025.

\bibitem{3GPP2019Study}
3GPP, \emph{{Study on channel model for frequencies from 0.5 to 100 GHz (Release 16)}}.\hskip 1em plus 0.5em minus 0.4em\relax Technical Report TR 38.901, available: http://www.3gpp.org, 2019.

\bibitem{vaswani2017attention}
A.~Vaswani, N.~Shazeer, N.~Parmar, J.~Uszkoreit, L.~Jones, A.~N. Gomez, L.~Kaiser, and I.~Polosukhin, ``{Attention is All you Need},'' in \emph{Advances in Neural Information Processing Systems 30: Annual Conference on Neural Information Processing Systems (NIPS)}, Long Beach, CA, USA, Dec. 2017, pp. 5998--6008.

\bibitem{radford2019language}
A.~Radford, J.~Wu, R.~Child, D.~Luan, D.~Amodei, I.~Sutskever \emph{et~al.}, ``{Language models are unsupervised multitask learners},'' \emph{OpenAI blog}, vol.~1, no.~8, p.~9, 2019.

\bibitem{Hu2021LoRALA}
E.~J. Hu, Y.~Shen, P.~Wallis, Z.~Allen-Zhu, Y.~Li, S.~Wang, L.~Wang, and W.~Chen, ``{Lora: Low-rank adaptation of large language models},'' \emph{arXiv preprint arXiv:2106.09685}, 2021.

\end{thebibliography}

\end{document}